\documentclass{article}
\usepackage{amsmath}
\usepackage[dvips]{graphics}
\usepackage{latexsym}
\usepackage{subfigure}
\DeclareFontFamily{OT1}{rsfs}{}
\DeclareFontShape{OT1}{rsfs}{m}{n}{ <-7> rsfs5 <7-10> rsfs7 <10->
; ; ; ; ; ; ; ; ; ; rsfs10}{}
\DeclareMathAlphabet{\mycal}{OT1}{rsfs}{m}{n}

\begin{document}
\newcommand{\bea}{\begin{eqnarray*}}
\newcommand{\eea}{\end{eqnarray*}}
\newcommand{\bean}{\begin{eqnarray}}
\newcommand{\eean}{\end{eqnarray}}
\newcommand{\eqs}[1]{Eqs. (\ref{#1})}
\newcommand{\eq}[1]{Eq. (\ref{#1})}
\newcommand{\meq}[1]{(\ref{#1})}
\newcommand{\fig}[1]{Fig. \ref{#1}}
\newcommand{\ppa}[2]{\left(\frac{\partial}{\partial #1}\right)^{#2}}
\newcommand{\pp}[2]{\left(\frac{\partial #1}{\partial #2}\right)}

\newcommand{\mm}{m_-}
\newcommand{\mpl}{m_+}
\newcommand{\km}{k_-}
\newcommand{\kp}{k_+}
\newcommand{\qp}{Q_+}
\newcommand{\qm}{Q_-}
\newcommand{\Bp}{B_+}
\newcommand{\Bm}{B_-}

\newcommand{\tri}{\delta}
\newcommand{\grad}{\nabla}
\newcommand{\pa}{\partial}
\newcommand{\pf}[2]{\frac{\pa #1}{\pa #2}}
\newcommand{\cla}{{\cal A}}
\newcommand{\aqt}{\frac{1}{4}\theta}

\newcommand{\rdt}{\dot r^2}
\newcommand{\eqn}{&=&}
\newcommand{\non}{\nonumber \\}

\title{\bf Destroying  extremal Kerr-Newman black holes with test particles}
\author{Sijie Gao\footnote{ Email: sijie@bnu.edu.cn} , Yuan Zhang\footnote{ Email: zhangyuan@mail.bnu.edu.cn}\\
Department of Physics, Beijing Normal University,\\
Beijing 100875, China}
\maketitle

\begin{abstract}
It has been shown that a nearly extremal black hole can be overcharged or overspun by a test particle if radiative and self-force effects are neglected, indicating that the cosmic censorship might fail. In contrast, the existing evidence in literature suggests that an extremal black hole cannot be overcharged or overspun in a similar process. In this paper, we show explicitly that even an exactly extremal black hole can be destroyed by a test particle, leading to a possible violation of the cosmic censorship. By considering higher-order terms, which were neglected in previous analysis, we show that the violation is generic for any extremal Kerr-Newman black hole with nonvanishing charge and angular momentum. We also find that the allowed parameter range for the particle is very narrow, indicating that radiative and self-force effects should be considered and may prevent violation of the cosmic censorship. \\


\end{abstract}

\section{Introduction}
If a singularity is not covered by a black hole horizon, it can be seen by distant observers and is called a naked singularity. The weak ``cosmic censorship'' conjecture states that naked singularities cannot be formed by gravitational collapse with physically reasonable matter \cite{penrose}. A precise statement of this conjecture was given in \cite{wald-cos}.  Although a general proof of this conjecture has not been given, evidence in favor of it has been found and discussed in the past few decades. One way of testing the cosmic censorship conjecture is to see whether the black hole horizon can be destroyed by an object falling into the black hole. In the seminal work, Wald \cite{wald72} proved that a test particle cannot destroy the horizon of an extremal Kerr-Newman black hole. This work has been revisited and extended by a number of authors in the last decade \cite{hubeny}-\cite{vitor4}. It is worth mentioning that gravitational lensing by naked singularities has been studied in the past decade \cite{vir1,vir2}, making observational test of the cosmic censorship possible.

  There were two crucial assumptions in Wald's treatment. First, the existing black hole is extremal. Second, only linear terms in the particle's energy, charge and angular momentum are kept in the analysis. By releasing the two assumptions, Hubeny showed that a nearly extremal Reissner-Nordstrom (RN) black hole can be overcharged by a test particle. Recently, Jacobson showed that a nearly extremal Kerr black hole can be overspun. These results apparently indicate violations of the cosmic censorship, at least, they point out that the test particle assumption may not be valid and the radiative and self-force effects should be considered.

 Note that the results in \cite{hubeny, ted} agree with Wald's in the extremal limit. So it seems that the cosmic censorship holds anyway when one tries to overcharge or overspin extremal black holes. However, an overlooked fact is that the authors of \cite{hubeny} and \cite{ted} only considered the RN black hole and Kerr black hole respectively, while Wald considered the combination, i.e., the Kerr-Newman (KN) black hole. To distinguish from RN and Kerr solutions, we shall refer to KN black holes as those with nonvanishing charge and angular momentum.  By reexamining Wald's arguments, we find that counter examples can be found if higher-order terms are included in the calculation (High-order terms have been considered in \cite{hubeny, ted} for RN and Kerr black holes, but caused no violation to the cosmic censorship in the extremal cases ). This tells us that the cosmic censorship is not safe even for extremal black holes. We further find that the allowed range of the particle's energy is very small, which means that the particle's parameters must be finely tuned. This suggests that radiative and self-force effects are necessary for a complete proof of the cosmic censorship. Although it is difficult to perform a full analysis on these effects, notable progress has been made recently. Barausse, Cardoso and Khanna \cite{vitor-prl, vitor-prd} showed that, for some orbits, the conservative self-force may have the right sign to prevent the violation of the cosmic censorship. Most recently, Zimmerman, Vega, Poisson and Haas \cite{poisson} incorporated the particle's electromagnetic self-force, and their numerical results have provided strong evidence supporting the cosmic censorship.

\section{Review of Wald's proof} \label{review}
In this section, we review the gedanken experiment in extremal charged Kerr black holes proposed by Wald\cite{wald72}. Consider the charged Kerr solution,
\bean
ds^2=g_{tt}dt^2+g_{rr}dr^2+g_{\theta\theta}d\theta^2+g_{\phi\phi}d\phi^2+2g_{t\phi}
dtd\phi\,.
\eean
Assume the vector potential is in the form,
\bean
A_{a}=A_t dt_a+A_\phi d\phi_a\,.
\eean
A charged particle with mass $m$ and charge $q$ moves in the spacetime with four-velocity
\bean
u^a=\dot t\ppa{t}{a}+\dot r\ppa{r}{a}+\dot \theta\ppa{\theta}{a}+\dot \phi\ppa{\phi}{a} \,.
\eean
The conserved energy and angular momentum are
\bean
E\eqn -t^a(m u_a+q A_a) \,,\label{ex}\\
L\eqn \phi^a(m u_a+q A_a) \,.\label{le}
\eean
Solving \eqs{ex} and \meq{le} for $\dot t$ and $\dot \phi$, we have
\bean
\dot t\eqn\frac{E g_{\phi\phi}+g_{t\phi}L+A_t g_{\phi\phi} q-A_\phi g_{t\phi} q}{
m(g_{t\phi}^2-g_{\phi\phi} g_{tt})} \,,\\
\dot\phi\eqn-\frac{E g_{t\phi}+g_{tt}L+A_t g_{t\phi} q-A_\phi g_{tt} q}{
m(g_{t\phi}^2-g_{\phi\phi} g_{tt})}\,.
\eean
Substituting the two formulas into
\bean
g_{ab}u^au^b=-1
\eean
and solving the quadratic equation for $E$, we find
\bean
E\eqn \frac{-g_{t\phi}L-q A_t g_{\phi\phi}+q A_\phi g_{t\phi}}{g_{\phi\phi}}\nonumber\\
&\pm& \frac{1}{g_{\phi\phi}}\sqrt{(g_{t\phi}^2-g_{\phi\phi}g_{tt})[L^2-2qLA_\phi+q^2 A_\phi^2+m^2g_{\phi\phi}(1+g_{rr}\dot r^2+g_{\theta\theta}\dot \theta^2)]} \non \,. \label{epm}
\eean
Note that $u^a$ is future pointing, which implies $\dot t>0$. Therefore, we should take the plus sign in front of the square root in \eq{epm}.
Consequently,
\bean
E\geq \frac{-g_{t\phi}L-q A_t g_{\phi\phi}+q A_\phi g_{t\phi}}{g_{\phi\phi}} \,.\label{egq}
\eean
The Kerr-Newmann metric is given by \cite{waldbook},
\bean
g_{tt}\eqn -\frac{\Delta-a^2 \sin^2\theta}{\Sigma} \,,\\
g_{t\phi}\eqn-\frac{a\sin^2\theta(r^2+a^2-\Delta)}{\Sigma}\,,\\
g_{\phi\phi}\eqn\frac{(r^2+a^2)^2-\Delta a^2\sin^2\theta}{\Sigma}\sin^2\theta \,,\\
A_t\eqn-\frac{Qr}{\Sigma},\ \ \ A_\phi=\frac{Qr}{\Sigma}a\sin^2\theta \,,\\
g_{rr}\eqn\frac{\Sigma}{\Delta} \,,\\
g_{\theta\theta}\eqn\Sigma\,,
\eean
with
\bean
\Sigma\eqn r^2+a^2\cos^2\theta \,,\\
\Delta\eqn r^2+a^2+Q^2-2Mr \,.
\eean
Then at the horizon $r=r_+$, \eq{epm} is written as
\bean
E=\frac{aL+qQr}{a^2+r^2}+m\sqrt{\frac{(a^2+2r_+^2+a^2\cos^2(2\theta))^2}{4(a^2+r_+^2)^2}\dot r^2}\label{ees} \,,
\eean
and thus
\bean
E\geq \frac{aL+qQr}{a^2+r^2}\,.
\eean
For an extremal black hole $r_+=M$, we have
\bean
E\geq \frac{aL+qQM}{a^2+M^2} \label{e1} \,.
\eean

On the other hand, to destroy the black hole horizon with $M^2=Q^2+a^2$, the particle must satisfy
\bean
(E+M)^2<(Q+q)^2+\left(\frac{aM+L}{M+E}\right) ^2 \label{e2}\,.
\eean
Expanding the last term around $E=0$, we have
\bean
E^2+M^2+2ME< Q^2+q^2+2qQ-\frac{(L+aM)^2}{M^2}+\frac{2(L+aM)^2E}{M^3}\,.
\eean
Using $M^2=Q^2+a^2$ and keeping the terms linear to $q,E,L$, we have
\bean
E<\frac{aL+MqQ}{M^2+a^2}\,,
\eean
which contradicts \eq{e1}. Thus, the cosmic censorship is upheld if higher-order terms are neglected. In the next section, we shall see that higher-order terms do not change this result if one attempts to destroy an extremal Kerr or RN black hole.

\section{Kerr and RN cases}
The above result is derived from a Kerr-Newman black hole. Now let us consider the following two reduced cases:

1. Pure Kerr ($Q=q=0, M=a$)

\eq{e1} reduces to
\bean
E\geq \frac{L}{2M}\,,
\eean
and \eq{e2} reduces to
\bean
E+M<\frac{M^2+L}{E+M}\,,
\eean
i.e.,
\bean
E^2+2ME<L\,,
\eean
\bean
E<\frac{L}{2M}-\frac{E^2}{2M}<\frac{L}{2M}\,,
\eean
so no solution can be found.

2. Pure RN($a=L=0, M=Q$)

\eq{e1} reduces to
\bean
E\geq q\,,
\eean
and \eq{e2} reduces to
\bean
E+M<Q+q \,,
\eean
i.e.,
\bean
E<q \,.
\eean
Obviously, there is no solution.

Thus, there is no violation of cosmic censorship for either Kerr black hole or RN black hole, agreeing with the results of Hubeny, Jacobson and Sotiriou \cite{hubeny,ted} . Differing from the treatment in Section \ref{review}, no linear approximation has been made in the above proof.

\section{Violation of the cosmic censorship for extremal KN black holes}
 From the last section, we see that the cosmic censorship conjecture has passed the test of gedanken experiments in extremal RN or  Kerr black holes, even without linear approximation. However, it is unknown whether higher-order terms can lead to a different conclusion for extremal KN black holes ($Q\neq 0$ and $a\neq 0$). We first show that the two inequalities \meq{e1} and \meq{e2} can be simplified and combined into one.
Define
\bean
W=(M+E)^2\,,
\eean
and rewrite \eq{e2} as
\bean
W^2-(Q+q)^2W-(aM+L)^2<0 \,.
\eean
This means
\bean
W_1<W<W_2\
\eean
with
\bean
W_{1,2}=\frac{(Q+q)^2\pm\sqrt{(Q+q)^4+4(aM+L)^2}}{2} \label{w12} \,.
\eean
From \eq{e1} we have
\bean
W>\left(\frac{aL+qQM}{a^2+M^2}+M\right)^2\equiv W_3 \,.
\eean
Obviously, $W_1<0$ and $W_2, W_3>0$.
Therefore, the necessary and sufficient condition for both inequalities \meq{e1} and \meq{e2} being satisfied is
\bean
W_2>W_3 \,,
\eean
i.e.,
\bean
s&\equiv& W_2-W_3 \\
&=& \frac{(Q+q)^2+\sqrt{(Q+q)^4+4(aM+L)^2}}{2}-\left(\frac{aL+qQM}{a^2+M^2}+M\right)^2
\label{s2} \\
&>&0 \label{s} \,.
\eean
Expanding \eq{s2} out to the second order in $q$ and $L$, we find
\bean
\frac{2a^2M^2(3M^2-a^2)}{(a^2+M^2)^3}q^2+\frac{M^2(-3a^2+M^2)}{(a^2+M^2)^3}L^2-\frac{2aM
Q(3M^2-a^2)}{(a^2+M^2)^3}qL>0 \,.\label{wql}
\eean

Now we can estimate the allowed range of $E$. From
\bean
W_3<W<W_2 \,,
\eean
we see the allowed range of $E$, denoted by $\Delta E$, satisfies
$2M\Delta E \sim W_2-W_3 $. Then \eq{wql} suggests that $\Delta E$ is of order
$q^2/M$ or $L^2/M^3$.

\begin{figure}[htmb]
\centering \scalebox{0.5} {\includegraphics{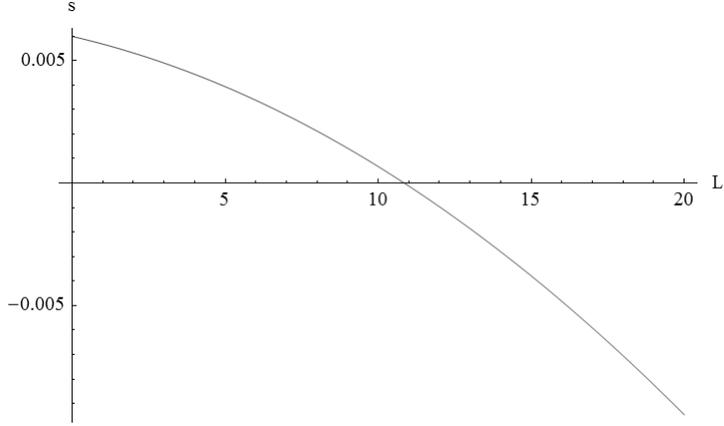}}
\caption{ Plot of $s-L$. For small values of $L$, $s$ is always positive.} \label{fig-sL}
\end{figure}

Note that the first term in \eq{wql} is always positive since $M^2\geq a^2$ for a KN black hole. So
\eq{wql} shows that as long as $Q\neq0$, $a\neq 0$ and $q\neq 0$, there  always exist solutions if $L$ is sufficiently small. To be specific, we choose the parameter set to be $M=100$ , $a=90$, and then $Q=\sqrt{M^2-a^2}=43.6$. We further choose $q=0.1$ such that the test body condition $q\ll Q$ is met. Now  $s$ in \eq{s} can be treated as a function of $L$. The plot in \fig{fig-sL} confirms that small values of $L$ always lead to positive $s$.

For illustration, we take $L=5$ and find  $4.8944\times 10^{-2}<E<4.8964\times 10^{-2}$. So $\Delta E\sim 2\times 10^{-5}$, which is comparable to $q^2/M=10^{-4}$ and $L^2/M^3=2.5\times 10^{-5}$, as expected.

\begin{figure}[htmb]
\centering \scalebox{0.5} {\includegraphics{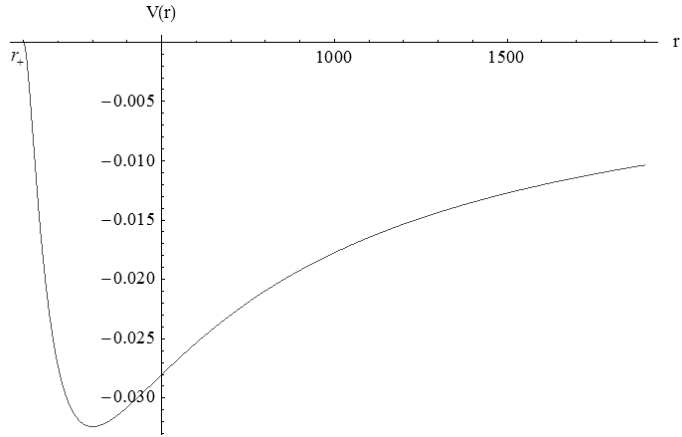}}
\caption{The effective potential is negative for all $r>r_+$. } \label{vr}
\end{figure}

Next, we show that such a particle can be released from infinity and falls all the way into the black hole. Since the metric is axisymmetric, there exist orbits lying entirely in the equatorial plane $\theta=\pi/2$. For such an orbit, one can solve \eq{epm} for $\dot r^2$ and obtain
\bean
\dot r^2=-V(r) \,,
\eean
where the effective potential $V(r)$ is given by
\bean
V(r)\eqn -\frac{1}{m^2 r^4} (a^4 E^2 - 2 a^3 E L + q^2 Q^2 r^2 - 2 E q Q r^3 +
   E^2 r^4 - L^2 \Delta  - m^2 r^2 \Delta \non
 &+&
   2 a L (q Q r + E (-r^2 + \Delta)) +
   a^2 (L^2 + E (-2 q Q r + 2 E r^2 - E \Delta)))\,.\non
   \label{vvr}
\eean
We still choose $M=100, a=90,q=0.1, L=5$ as above, and $m=E=0.048955$ such that $E$ is in the allowed range. Numerical calculation shows that $V(r)$ is negative for all $r\geq r_+$ (see \fig{vr}). It is easy to check that our choice $m=E$ indicates that the particle stays at rest relative to a stationary observer at infinity, so this initial condition is realizable in practice.

\section{Discussion and Conclusions}
We have shown that, without taking into account the radiative and self-force effects, a test particle may destroy the horizon of an extremal charged Kerr black hole, resulting in an apparent violation of the cosmic censorship. The violation is generic for any extremal KN black hole. As shown by Wald \cite{wald72}, there would be no violation if higher-order terms are neglected.  We also show that the energy of the particle must be finely tuned, i.e., the allowed range of energy $\Delta E$ is of order $q^2/M$ or $L^2/M^3$. A similar fine tuning has been pointed out and discussed in \cite{ted2} for nearly extremal Kerr black holes.  Smith and Will \cite{charge-sch} show that a charged particle in Schwarzschild spacetime will feel a repulsive electrostatic self-force induced by the spacetime curvature. Consequently, the particle has an additional self-interacting energy with magnitude $Mq^2/r^2$ \cite{hod}. If we use this result to estimate the magnitude of the self-force correction to the energy of a particle outside a RN black hole, it becomes $q^2/M$ at the extremal black horizon $r=M$, which is the same order as $\Delta E$ we discussed above.  This indicates that the self-force effect is important in testing the cosmic censorship. Despite the self-force effect, there is another open issue related to this
scenario. A hidden assumption in the above argument is that once the black hole absorbs the particle, it will settle down to a new stationary state. However, this result is not guaranteed by current theories \cite{ted2}. So far, all results can only be taken as some indication that cosmic censorship might fail.

\section*{Acknowledgements}
This research was supported by NSFC Grants No. 10605006, 10975016 and 11235003.

\end{document}